\documentclass[aps,pra,showpacs,twocolumn]{revtex4-1}
\usepackage{amsfonts}
\usepackage{amssymb}
\usepackage{amsmath}
\usepackage{graphicx}
\usepackage{epsfig}

\begin{document}

\title{Perfect State Transfer in $\mathcal{PT}$-symmetric Non-Hermitian
Networks}
\author{X. Z. Zhang, L. Jin and Z. Song}
\affiliation{School of Physics, Nankai University, Tianjin 300071, China}
\email{songtc@nankai.edu.cn}

\begin{abstract}
We systematically study the $\mathcal{PT}$ (parity-time reversal)-symmetric
non-Hermitian version of a quantum network proposed in the work of
Christandl \textit{et al.} [Phys. Rev. Lett. 92, 187902 (2004)]. The
exclusive nature of this model show that it is a nice paradigm to
demonstrate the complex quantum mechanics theory for the relationship
between the pseudo-Hermitian Hamiltonian and its Hermitian counterpart, as
well as a candidate in experimental realization to simulate the $\mathcal{PT}
$ symmetry breaking. We also show that this model allows conditional perfect
state transfer within the unbroken $\mathcal{PT}$ symmetry region, but not
arbitrary. This is due to the fact that the evolution operator at certain
period is equivalent to the $\mathcal{PT}$ operator for the real-valued
wavefunction in the elaborate $\mathcal{PT}$-symmetric Hilbert space.
\end{abstract}

\pacs{11.30.Er, 03.67.Hk, 03.65.-w}
\maketitle


\section{Introduction}

\label{sec_intro}The transmission of quantum state through a solid state
data bus with minimal spatial and dynamical control is an experimental
challenging and a theoretically necessary task for implementing a scalable
quantum computation based on realistic silicon devices. S. Bose \cite{S.
Bose} first demonstrated the possibility that in a solid state based\
quantum computer, local interactions can apply entangling gates between
distant qubits. In principle, perfect transfer of quantum state can be
implemented by specifically engineering chain \cite{M. Christandl1,M.
Christandl2,T.Shi,Xiao-Feng Qian}. It is also showed that a quantum system
possessing a commensurate structure of energy spectrum matched with the
corresponding symmetry can ensure the perfect quantum state transfer \cite%
{M. Christandl2,M. Christandl2,T.Shi,Y.Li}.

The aim of this paper is to extend these findings to the non-Hermitian
system. It is motivated by the interest in complex potentials in both
theoretical and experimental aspects. Much effort has been devoted to
establish a parity-time ($\mathcal{PT}$) symmetric quantum theory as a
complex extension of the conventional quantum mechanics \cite{Bender
98,Bender 99,Dorey 01,Bender 02,A.M43,A.M,A.M36,Jones} since the seminal
discovery by Bender \cite{Bender 98}. It is found that the non-Hermitian
Hamiltonian with simultaneous $\mathcal{PT}$ symmetry can have an entirely
real quantum mechanical energy spectrum and has profound theoretical and
methodological implications. Researches and findings relevant to the spectra
of the $\mathcal{PT}$-symmetric systems were presented, such as exceptional
points \cite{EP}, spectral singularities for complex scattering potentials
\cite{AMSS},\ as well as\emph{\ }complex crystal and other specific models
\cite{LonghiSS}. Furthermore, quantum dynamics in open systems and\emph{\ }$%
\mathcal{PT}$-symmetric non-Hermitian systems were investigated \cite%
{Graefe,Kottos, LJin84}. At the same time the $\mathcal{PT}$\ symmetry is
also of great relevance to the technological applications based on the fact
that the imaginary potential could be realized by complex index in optics
\cite{Bendix,Joglekar,Keya,YDChong,LonghiLaser}.\ In fact, such $\mathcal{PT}
$ optical potentials can be realized through a judicious inclusion of index
guiding and gain/loss regions, and the most interesting aspects associated
with $\mathcal{PT}$-symmetric system were observed during the dynamic
evolution process \cite{Klaiman,El-Ganainy,Makris,Musslimani}.

According to the complex quantum mechanics, a non-Hermitian Hamiltonian with
entire real spectrum can be transformed into a Hermitian Hamiltonian via the
well-established metric operator formalism \cite{A.M37}.\ Nevertheless, the
example is rare in which explicit derivations can be performed to evaluate
the equivalent Hermitian counterpart.

In this paper, we systematically study the $\mathcal{PT}$ (parity-time
reversal)-symmetric non-Hermitian version of a quantum network proposed in
the work of Christandl \textit{et al.} \cite{M. Christandl1,M. Christandl2}%
.\ The main results are as follows: (i) The non-Hermitian version of a
quantum network is exact solved and the critical behavior is analytically
studied at the exceptional point. (ii)\ We provide an explicit mapping
between the proposed pseudo-Hermitian Hamiltonian and its equivalent
Hermitian Hamiltonian. The present model is a nice paradigm to demonstrate
the relationship between the non-Hermitian Hamiltonian and its Hermitian
counterpart, since all of Hamiltonians with different parameters, no matter
Hermitian or non-Hermitian, are local and have the equally spaced spectra.
(iii) Furthermore, we investigate the connection between the present model
and the\textbf{\ }$\mathcal{PT}$-symmetric\ hypercube graph. (iv)\ We also
show\ that such an extended model still allows conditional perfect state
transfer within the unbroken $\mathcal{PT}$ symmetry region, but not
arbitrary. This is due to the fact that the evolution operator at certain
period is equivalent to the $\mathcal{PT}$ operator for the real-valued
wavefunction in the elaborate $\mathcal{PT}$-symmetric Hilbert space.

This paper is organized as follows. In Section \ref{sec_model}, we present
the model and the solutions. In Section \ref{sec_nonh}, we explore the basic
properties of the model in its non-Hermitian version. In Section \ref%
{sec_metric}, we investigate the metric and Hermitian counterpart for the
extended model. In Section \ref{sec_hypercube}, we investigate the
connection between the present model and the hypercube graph. Section \ref%
{sec_perfectstate} is devoted to the dynamics of the model.\ Finally, we
give a summary and discussion in Section \ref{sec_summary}.

\section{The model}

\label{sec_model}We start with the Hamiltonian for an $N$-site tight-binding
chain with linear potentials,

\begin{eqnarray}
H &=&\frac{1}{2}\sum\limits_{l=1}^{N-1}\sqrt{l\left( N-l\right) }\left(
a_{l}^{\dagger }a_{l+1}+\text{H.c.}\right)   \label{H} \\
&&+\sum\limits_{l=1}^{N}\left[ \frac{1}{2}\left( N+1\right) -l\right] \gamma
a_{l}^{\dagger }a_{l}  \notag
\end{eqnarray}%
where $a_{l}^{\dagger }$ ($a_{l}$) is the creation (annihilation) operator
at site $l$. For the sake of simplicity we take the units of coupling
constant as\textbf{\ }$1$. The conclusion of this paper is valid for both
fermion and boson systems. In the case of $\gamma =0$, it is reduced to the
model in Ref. \cite{M. Christandl1,M. Christandl2}, which has been shown
\cite{Y.Li} to guarantee that the Hamiltonian $H(\gamma =0)$ evolves states $%
\left\vert \phi \right\rangle $ into $\mathcal{P}\left\vert \phi
\right\rangle $ at the time $\pi $, no matter what these states are. Here
parity operator $\mathcal{P}$ is given by $\mathcal{P}a_{l}^{\dagger }%
\mathcal{P}^{-1}=a_{N+1-l}^{\dagger }$. This due to the fact that the
Hamiltonian $H(\gamma =0)$\ is $\mathcal{P}$-symmetric and possesses an
equally spaced spectrum \cite{T.Shi}. In the case of nonzero $\gamma $, the $%
\mathcal{P}$\ symmetry is broken. In the following, we will show that it
still has an equal-spaced spectrum. Here the idea is to treat the
Hamiltonian as an angular momentum in an external magnetic field.

Defining the operators\
\begin{eqnarray}
J^{+} &=&\left( J^{-}\right) ^{\dag }=\sum\limits_{l=1}^{N-1}\sqrt{l\left(
N-l\right) }a_{l}^{\dagger }a_{l+1},  \label{J+-z} \\
J_{x} &=&\frac{J^{+}+J^{-}}{2},\text{ }J_{y}=\frac{J^{+}-J^{-}}{2i},  \notag
\\
J_{z} &=&\sum\limits_{l=1}^{N}\left( \frac{N+1}{2}-l\right) a_{l}^{\dagger
}a_{l},  \notag
\end{eqnarray}%
which satisfy the following angular momentum commutation relations%
\begin{equation}
\left[ J^{+},J^{-}\right] =2J_{z},\text{ }\left[ J^{z},J^{\pm }\right] =\pm
J^{\pm }.  \label{commutation}
\end{equation}%
Then $J_{x,y,z}$\ acts as the angular momentum operator, and the Hamiltonian
can be rewritten as%
\begin{equation}
H=J_{x}+\gamma J_{z}=\overrightarrow{J}\cdot \overrightarrow{B}  \label{JB}
\end{equation}%
where%
\begin{equation}
\overrightarrow{B}=\left( B_{x},B_{y},B_{z}\right) =\left( 1,0,\gamma
\right) .  \label{B}
\end{equation}%
Obviously, it can be diagonalized as%
\begin{equation}
H=\sqrt{1+\gamma ^{2}}J_{n}^{\prime }  \label{H'}
\end{equation}%
with $J_{n}^{\prime }$ being a $n$-component of angular momentum operator,
where $n=\overrightarrow{B}/\left\vert \overrightarrow{B}\right\vert $ is
the unit vector in the field direction. In this paper, we concentrate on the
single-particle invariant subspace, which corresponds to the angular
momentum system with $J$ $=(N-1)/2$. Then the energy levels are still
equally spaced.\ In this subspace, the eigenvector of $H$\ can be obtained
from that of $H(\gamma =0)$\ by the rotation operator, i.e.,%
\begin{eqnarray}
\left\vert \psi _{n}\right\rangle  &=&e^{-i\left[ \beta \left( \gamma
\right) -\pi /2\right] J_{y}}\left\vert \psi _{n}\left( \gamma =0\right)
\right\rangle   \label{psi_n1} \\
&&\left( n=1,2,...N\right) .  \notag
\end{eqnarray}%
Here $\beta \left( \gamma \right) $\ is an angle
\begin{equation}
\beta \left( \gamma \right) =\arctan \left( \frac{1}{\gamma }\right)
\label{beta}
\end{equation}%
and $\left\vert \psi _{n}\left( \gamma =0\right) \right\rangle $ is the
eigenvector of $H(\gamma =0)$, i.e.,%
\begin{equation}
H(\gamma =0)\left\vert \psi _{n}\left( \gamma =0\right) \right\rangle
=\left( \frac{N+1}{2}-n\right) \left\vert \psi _{n}\left( \gamma =0\right)
\right\rangle   \label{H_gamma0}
\end{equation}%
which can be further expressed as%
\begin{equation}
\left\vert \psi _{n}\left( \gamma =0\right) \right\rangle
=\sum\limits_{l}d_{n,l}\left( \frac{\pi }{2}\right) a_{l}^{\dagger
}\left\vert 0\right\rangle .  \label{psi_gamma0}
\end{equation}%
Then we have
\begin{equation}
\left\vert \psi _{n}\right\rangle =\sum\limits_{l}d_{n,l}\left( \beta
\right) a_{l}^{\dagger }\left\vert 0\right\rangle   \label{psi_n2}
\end{equation}%
where $d_{n,l}\left( \beta \right) $ is Winger $d$-functions%
\begin{eqnarray}
&&d_{n,l}\left( \beta \right) =d_{n,l}^{\left[ \left( N-1\right) /2\right]
}\left( \beta \right)   \label{d matrix} \\
&=&\left[ \left( N-n\right) !\left( n-1\right) !\left( N-l\right) !\left(
l-1\right) !\right] ^{1/2}  \notag \\
&&\times \sum\limits_{\nu }\frac{\left( -1\right) ^{\nu }\left( \cos \frac{%
\beta }{2}\right) ^{N-1+l-n-2\nu }\left( -\sin \frac{\beta }{2}\right)
^{n-l+2\nu }}{\left( l-1-\nu \right) !\left( N-n-\nu \right) !\left( \nu
+n-l\right) !\nu !}.  \notag
\end{eqnarray}%
We define $\psi _{n}\left( l\right) =$\ $d_{n,l}\left( \beta \right) $\ to
express the orthonormal relation as%
\begin{equation}
\sum_{l}\psi _{m}\left( l\right) \psi _{n}\left( l\right) =\delta _{mn}.
\label{orthnormal}
\end{equation}%
It is important to note that the above relation is still true for imaginary $%
\gamma $ except the points $\gamma =\pm i$.

%
%
%
%
%

\section{Non-Hermitian $\mathcal{PT}$-symmetric Hamiltonian}

\label{sec_nonh}Now we consider the Hamiltonian of Eq. (\ref{H}) with
imaginary linear potentials by taking $\gamma \longrightarrow i\gamma $,
which can be written as%
\begin{eqnarray}
\mathcal{H} &=&\frac{1}{2}\sum\limits_{l=1}^{N-1}\sqrt{l\left( N-l\right) }%
\left( a_{l}^{\dagger }a_{l+1}+\text{H.c.}\right)  \label{H_i} \\
&&+i\gamma \sum\limits_{l=1}^{N}\left[ \frac{1}{2}\left( N+1\right) -l\right]
a_{l}^{\dagger }a_{l}.  \notag
\end{eqnarray}%
Mathematically, all the solutions of $H$ can be extended to that of $%
\mathcal{H}$ by simply taking $\gamma \longrightarrow i\gamma $ except for
the points $\left\vert \gamma \right\vert =1$, since it does not induce any
singularity in the rotation operator $e^{-i\beta \left( i\gamma \right)
J_{y}}$.\ This fact accords with complex quantum mechanics. Note that the
non-Hermitian Hamiltonian $\mathcal{H}$\ is $\mathcal{PT}$-symmetric, i.e., $%
\left[ \mathcal{PT},\mathcal{H}\right] =0$. The antilinear time-reversal
operator is defined as $\mathcal{T}i\mathcal{T}^{-1}=-i$. The phase diagram
of this system is determined by critical (exceptional) points\ $\left\vert
\gamma \right\vert =1$. This model exhibits two phases: an unbroken symmetry
phase with a purely real energy spectrum
\begin{equation}
\varepsilon _{n}=\sqrt{1-\gamma ^{2}}\left( \frac{N+1}{2}-n\right)
,n=1,2,...,N  \label{ep_k}
\end{equation}%
when the potentials are in the region $\left\vert \gamma \right\vert <1$ and
a spontaneously broken symmetry phase with an imaginary spectrum when the
potentials are in the region $\left\vert \gamma \right\vert >1$. In this
paper, we only focus on the $\mathcal{PT}$-symmetric\ region. Although the
spectrum is real, the corresponding eigenfunctions are no longer orthonormal
with respect to the Dirac inner product. One can establish the complete
biorthogonal set by the eigenfunctions of the Hamiltonian $\mathcal{H}%
^{\dagger }$. Denoting
\begin{equation}
\left\vert \phi _{n}\right\rangle =\sum\limits_{l}\phi _{n}\left( l\right)
a_{l}^{\dagger }\left\vert 0\right\rangle =\sum\limits_{l}\psi _{n}^{\ast
}\left( l\right) a_{l}^{\dagger }\left\vert 0\right\rangle ,
\end{equation}%
we have
\begin{equation}
\mathcal{H}^{\dagger }\left\vert \phi _{n}\right\rangle =\varepsilon
_{n}\left\vert \phi _{n}\right\rangle .
\end{equation}%
Then Eq. (\ref{orthnormal}) leads to the orthonormal relation

\begin{equation}
\left\langle \phi _{m}\right. \left\vert \psi _{n}\right\rangle
=\sum_{l}\phi _{m}^{\ast }\left( l\right) \psi _{n}\left( l\right) =\delta
_{mn}.  \label{complete}
\end{equation}%
Furthermore, the complete set obeys the relation%
\begin{equation}
\mathcal{PT}\left\vert \psi _{n}\right\rangle =(-1)^{n}\left\vert \psi
_{n}\right\rangle .  \label{PT}
\end{equation}%
In the case of $\gamma =0$, the above relation is reduced to
\begin{equation}
\mathcal{P}\left\vert \psi _{n}\left( \gamma =0\right) \right\rangle
=(-1)^{n}\left\vert \psi _{n}\left( \gamma =0\right) \right\rangle .
\label{PT_gamma0}
\end{equation}%
It is worth to note that Eqs. (\ref{PT}) and (\ref{PT_gamma0}) have
different implications: If one obtains a set of eigenfunctions satisfying
Eq. (\ref{H_gamma0}), they will obey Eq. (\ref{PT_gamma0}) spontaneously.
However, Eq. (\ref{PT}) is not necessary for the eigenfunctions of the
Hamiltonian $\mathcal{H}(\gamma )$. This is due to the fact that operator $%
\mathcal{T}$\ is a antilinear operator. This issue will be elaborated in the
next section when we investigate the application of Eq. (\ref{PT})
associated with the dynamic process.

Now we consider the case of $\left\vert \gamma \right\vert =1$, which are
singular points for $\psi _{n}\left( l\right) $. However, one can construct
the eigenfunctions of the Hamiltonians $\mathcal{H}_{\pm }=\mathcal{H}\left(
\gamma =\pm 1\right) $ as the form
\begin{eqnarray}
\left\vert \varphi _{\pm }\right\rangle  &=&\sum\limits_{l}\varphi _{\pm
}\left( l\right) a_{l}^{\dagger }\left\vert 0\right\rangle   \notag \\
&=&\sum\limits_{l}\left( \sqrt{2}\right) ^{1-N}\sqrt{C_{N-1}^{l-1}}\left(
\pm i\right) ^{n-l}a_{l}^{\dagger }\left\vert 0\right\rangle ,  \label{phi+-}
\end{eqnarray}%
where\textbf{\ }%
\begin{equation}
\varphi _{\pm }\left( l\right) =\left( \sqrt{2}\right) ^{1-N}\sqrt{%
C_{N-1}^{l-1}}\left( \pm i\right) ^{n-l}
\end{equation}%
Acting the Hamiltonians $\mathcal{H}_{\pm }$ on the states $\left\vert
\varphi _{\pm }\right\rangle $, straightforward algebra shows that
\begin{equation}
\mathcal{H}_{\pm }\left\vert \varphi _{\pm }\right\rangle =0.
\end{equation}%
Eigenstates $\left\vert \varphi _{\pm }\right\rangle $\ are zero-norm
states, i.e.,%
\begin{equation}
\sum\limits_{l}\varphi _{\pm }^{2}\left( l\right) =0.
\end{equation}%
%
%
%
%
%
%
%
%
%
%
%
%
%
%
%
%
%
%
%
%
%
%
%
%
%
%
%
%
%
%
%
%
%
%
%
%
%
%
%
%
%
%
%
%
%
%
%
%
%
%
%
%
%
%
%
%
%
%
%
%
%
%
%
%
%
%
%
%

\begin{figure}[tbp]
\includegraphics[ bb=37 168 500 600, width=8.5 cm, clip]{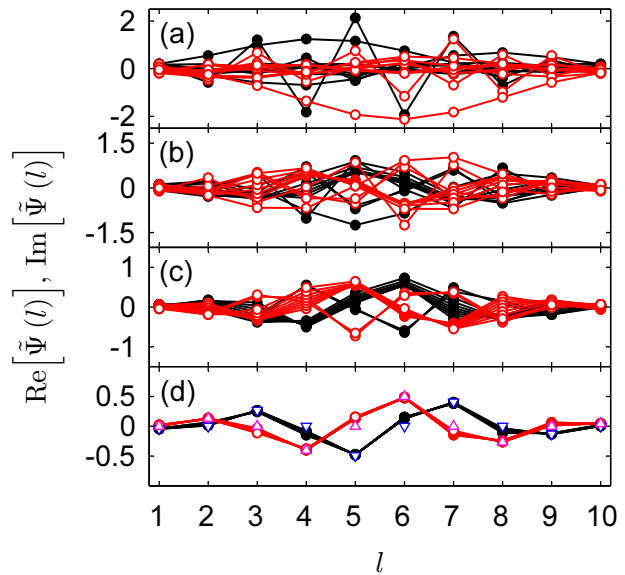}
\caption{(Color online) Plots of the dressed eigenfuctions as functions of \
for $N=10$ chain. The solid (empty) circle indicates the real (imaginary)
part of the wavefunction for (a) $\protect\gamma =0.2$, (b) $\protect\gamma %
=0.7$, (c) $\protect\gamma =0.9$, and (d) $\protect\gamma =0.998$,
respectively. It can be observed that as $\protect\gamma $\ tends to $1$,
all the wavefunctions approach to the same function Eq. (\protect\ref{phi+-}%
) for $N=10$, which is plotted in (d) (triangles). Note that the scales of
the subfigures for various $\protect\gamma $\ are different.}
\label{fig1}
\end{figure}


We can see that $\left\vert \gamma \right\vert =1$ is the boundary of two
phases, possessing the characteristics of exceptional point: The spectrum
exhibits square-root-type level repulsion in the vicinity of $\left\vert
\gamma \right\vert =1$. On the other hand, states $\left\vert \varphi _{\pm
}\right\rangle $ should be the coalescence of the\textbf{\ }eigenfunctions $%
\left\{ |\psi _{n}\rangle \right\} $\ as $\left\vert \gamma \right\vert
\rightarrow 1$. To demonstrate this point, we introduce the dressed
eigenfunctions $\{|\tilde{\psi}_{n}\rangle \}$\ by multiplying factor $%
g_{n}\left( \gamma \right) $\ to the original eigenfunctions $\left\{ |\psi
_{n}\rangle \right\} $,
\begin{equation}
\tilde{\psi}_{n}\left( l\right) =g_{n}\left( \gamma \right) \psi _{n}\left(
l\right)   \label{dressed}
\end{equation}%
where%
\begin{equation}
g_{n}\left( \gamma \right) =\left( \sqrt{2}\cos \frac{\beta }{2}\right)
^{1-N}\left( C_{N-1}^{n-1}\right) ^{-\frac{1}{2}}.  \label{factor}
\end{equation}%
We notice that%
\begin{equation}
\lim_{\pm \gamma \rightarrow 1}\tilde{\psi}_{n}\left( l\right) =\varphi
_{\pm }\left( l\right) ,  \label{lim}
\end{equation}%
i.e., all the dressed eigenstates coalesce with\ $\left\vert \varphi _{\pm
}\right\rangle $\ at the critical point. For illustration, numerical
simulation of finite size chains with various values of $\gamma $ is
performed. In Fig. \ref{fig1} we plot the dressed eigenfuctions including
real and imaginary parts for $N=10$ chain, respectively. It shows that all
the eigenfunctions tend to $\left\vert \varphi _{\pm }\right\rangle $\ as $%
\gamma $\ approaches the critical point.

Finally, we would like to point out that the nature of the\textbf{\ }$%
\mathcal{PT}$ symmetry breaking in such a model is exclusive.\ The entire
spectrum becomes imaginary and the $\mathcal{PT}$ symmetry of all the
eigenstates is broken at the point $\left\vert \gamma \right\vert =1$\
simultaneously. The degrees of $\mathcal{PT}$\ symmetry breaking is defined
as the fraction of eigenvalues that become complex \cite{scott}, the degree
of our system is $1$. This feature should lead to clear signatures in the
dynamics of the wavepacket. Such a model is a good candidate to simulate the
critical behavior in experiments.

\section{Metric and Hermitian counterpart}

\label{sec_metric}Another theoretical interest in the non-Hermitian $%
\mathcal{PT}$-symmetric system is the physical meaning of the Hamiltonian $%
\mathcal{H}$. When speaking of the physical significance of a non-Hermitian
Hamiltonian, one of the ways is to seek its Hermitian counterparts \cite%
{A.M38,A.M39,A.M13495} possessing the same real\ spectrum. According to the
complex quantum mechanics, a non-Hermitian $\mathcal{PT}$-symmetric
Hamiltonian can be transformed into a Hermitian Hamiltonian. This is
achieved by introducing a metric, a bounded positive-definite Hermitian
operator $\eta $, which can be constructed via the eigenstates of $\mathcal{H%
}^{\dag }$ \cite{A.M37}. However, the obtained equivalent Hermitian
Hamiltonian is usually quite complicated \cite{A.M37,J.L}. It is tough to
provide an explicit mapping of a pseudo-Hermitian Hamiltonian to its
equivalent Hermitian Hamiltonian. Fortunately, the present model is a nice
paradigm to demonstrate the relationship between the pseudo-Hermitian
Hamiltonian and its Hermitian counterpart. In this section, we will
illustrate the basic ideas via the above analytically solution.

Consider the single-particle case, the metric operator $\eta $\ can be
constructed via the eigenstates of $H^{\dag }$\ as

\begin{eqnarray}
\eta &=&\sum_{n}\left\vert \phi _{n}\right\rangle \left\langle \phi
_{n}\right\vert  \label{metric} \\
&=&\sum\limits_{n,l,l^{\prime }}d_{n,l^{\prime }}\left( \beta \right)
d_{n,l}^{\ast }\left( \beta \right) a_{l}^{\dagger }\left\vert
0\right\rangle \left\langle 0\right\vert a_{l^{\prime }}  \notag \\
&=&e^{-i\left( \beta ^{\ast }-\beta \right) J_{y}},  \notag
\end{eqnarray}%
which guarantees the relation%
\begin{equation}
\eta \mathcal{H}\eta ^{-1}=\mathcal{H}^{\dag }.  \label{pseudo}
\end{equation}%
It is noticed that the matrix representation of $\eta \mathcal{H}$\ and $%
\eta $ based on the orthonormal basis, say $\{a_{l}^{\dagger }\left\vert
0\right\rangle \}$\ under the Dirac inner product, are Hermitian matrices.
Operator $\eta $\ is called the metric operator since it can be used to
define the biorthogonal inner product, under which the unitary evolution can
be obtained. Furthermore, let
\begin{equation}
\rho =\sqrt{\eta }=e^{-\frac{i}{2}\left( \beta ^{\ast }-\beta \right) J_{y}}
\label{rol}
\end{equation}%
be the unique positive-definite square root of $\eta $. Then the Hermitian
operator $\rho $ acts as a similarity transformation\ to map the
non-Hermitian Hamiltonian $\mathcal{H}$\ onto its equivalent Hermitian
counterpart $h$ by

\begin{eqnarray}
h &=&\rho \mathcal{H}\rho ^{-1}  \label{h} \\
&=&\sqrt{\left( 1-\gamma ^{2}\right) }J_{x}  \notag \\
&=&\sqrt{\left( 1-\gamma ^{2}\right) }H(\gamma =0).  \notag
\end{eqnarray}%
It can be regarded as the Hermitian counterpart of the non-Hermitian
Hamiltonian $H(\gamma )$. It is fortunate that both the non-Hermitian and
the Hermitian Hamiltonian have simple structure: they possess the localized
couplings and have equally spaced spectra. The physics of both the original
non-Hermitian Hamiltonian and its equivalent Hermitian counterpart are
clear: they can be regarded as either the lattice model with
nearest-neighbor couplings (engineered chain with imaginary linear on-site
potentials) or the angular momentum coupled to the external complex field.
Additionally, the physics of $\rho $\ is also obviously,\ which presents the
operation of two successive rotations, $e^{-i\left( \beta ^{\ast }/2\right)
J_{y}}$\ and $e^{i\left( \beta /2\right) J_{y}}$, which denote the rotations
about $y$\ axis with the angle $\beta ^{\ast }/2$\ and\ $-\beta /2$,
respectively.

\begin{figure*}[tbp]
\includegraphics[ bb=23 75 550 750, width=5 cm, clip]{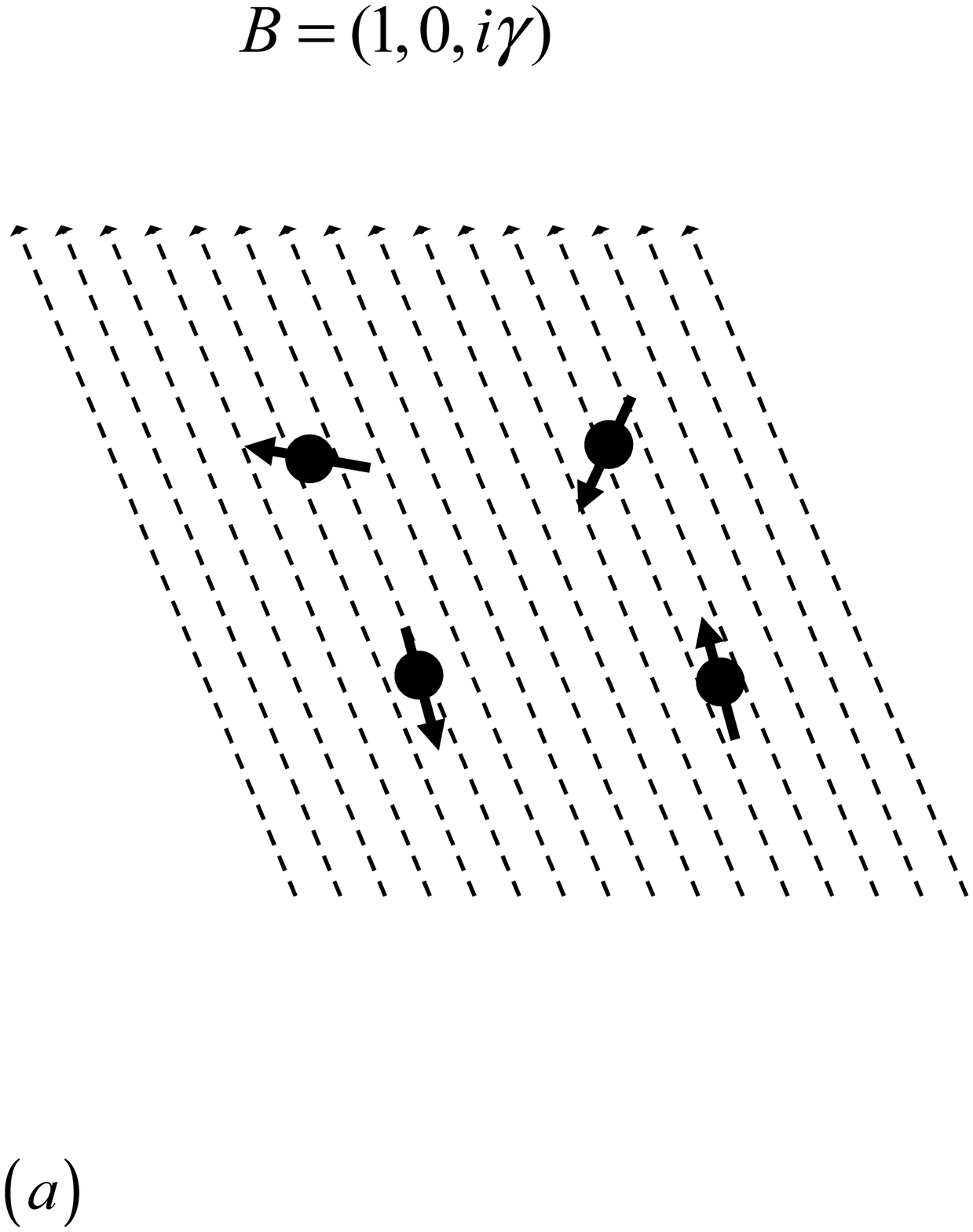} %
\includegraphics[ bb=23 75 560 720, width=5 cm, clip]{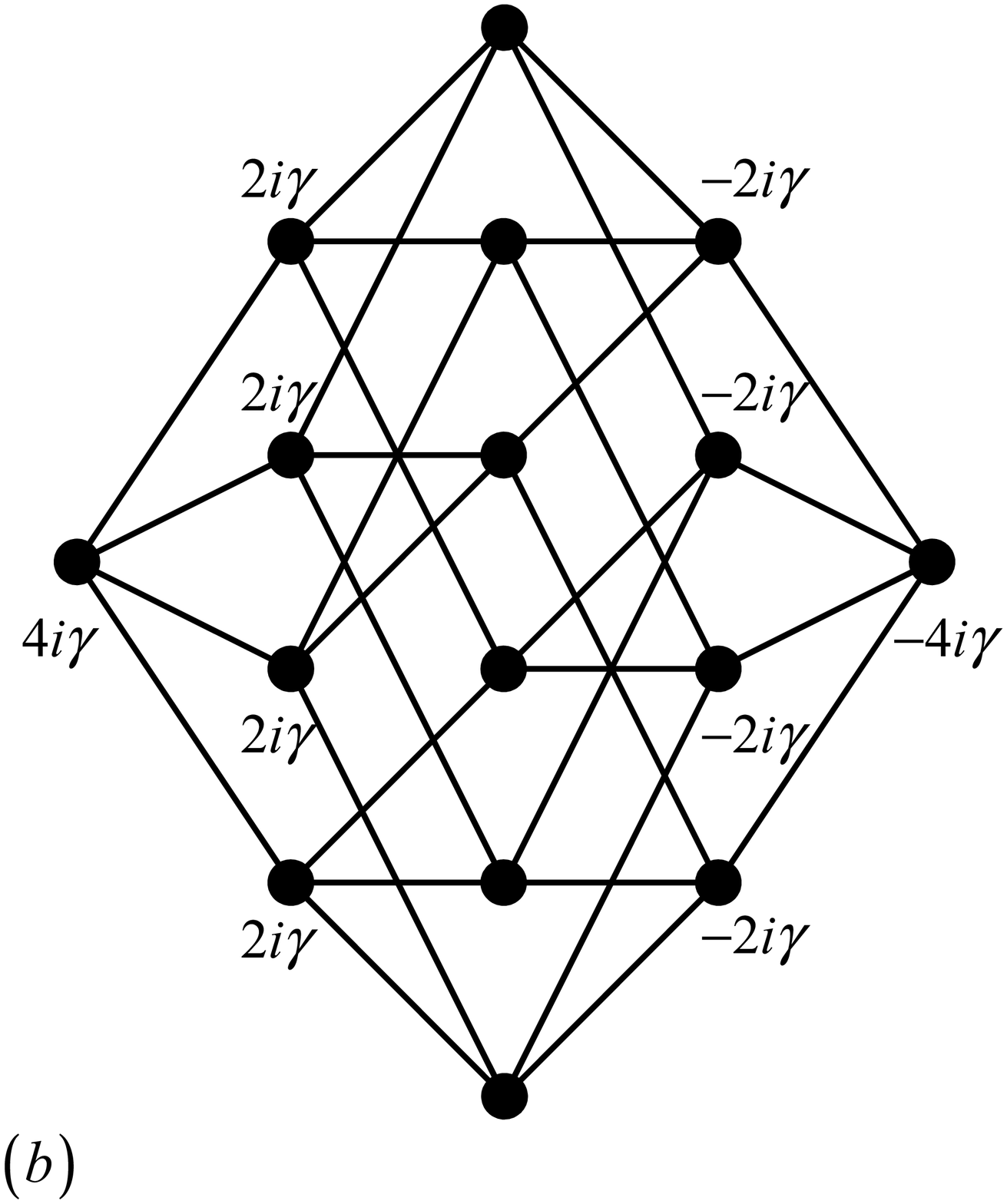} %
\includegraphics[ bb=23 75 549 741, width=5 cm, clip]{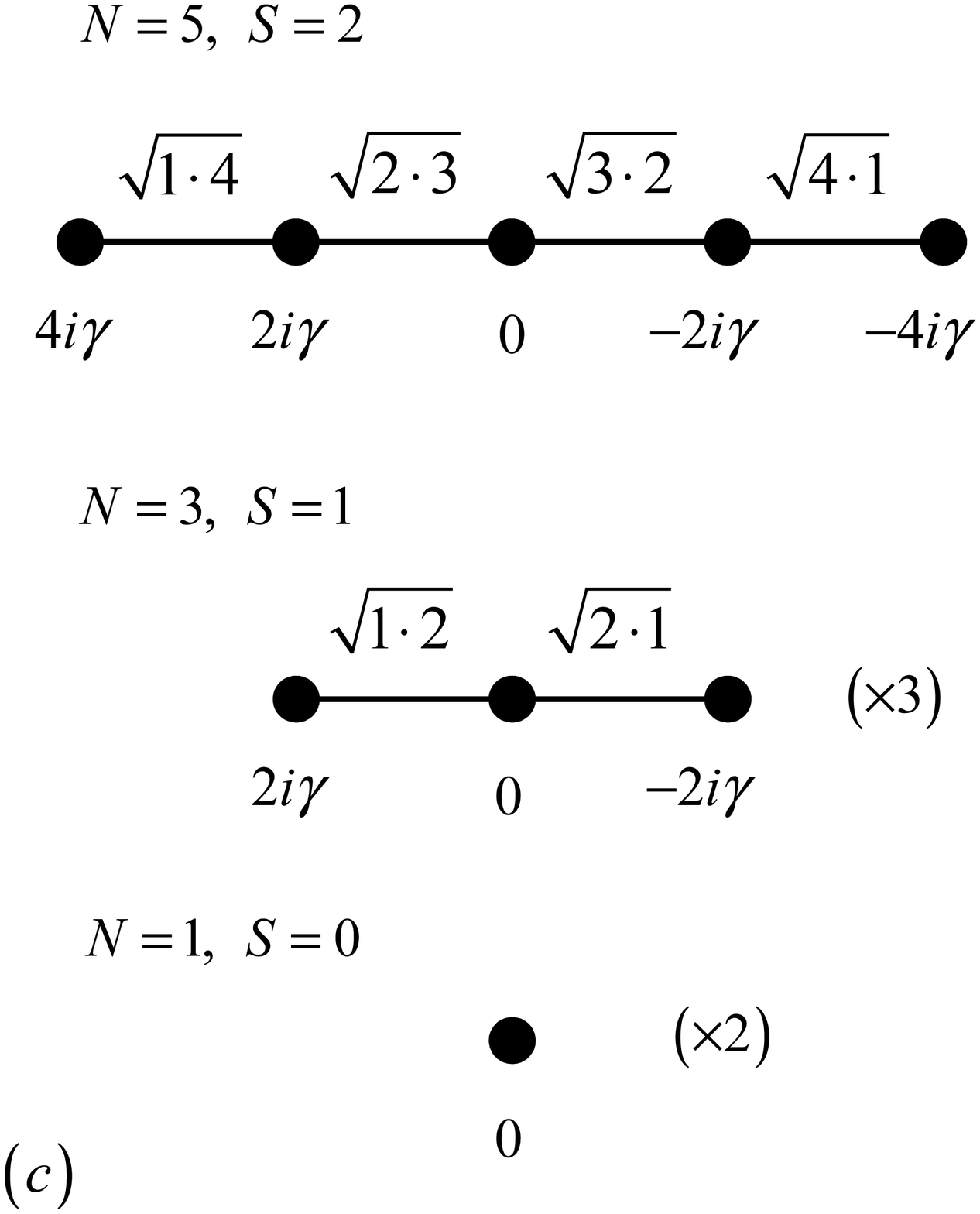}
\caption{(Color online) Schematics for the system with $d=4$ to illustrate
the connection among the systems of Eqs. (\protect\ref{H_i}), (\protect\ref%
{H_spin}), and the $\mathcal{PT}$-symmetric hypercube lattice. (a) Four
non-interacting spins in the external complex magnetic field. (b) The
five-column hypercube graph with imaginary potentials. (c) The projection of
the hypercube lattice onto the invariant subspaces denoted by $S$. In each
invariant subspace, the reduced Hamiltonians are described by Eq. (\protect
\ref{H_i})\ with different length, respectively.}
\label{fig2}
\end{figure*}


\section{$\mathcal{PT}$-symmetric hypercube}

\label{sec_hypercube}It has been pointed that the one-dimensional system $%
H(\gamma =0)$ can be achieved by the projection of the hypercube \cite{M.
Christandl2}. In this section, we will extend this approach to the
non-Hermitian regime and investigate the solutions for all the possible
projections of the hypercube.

We start the investigation with an ensemble of non-interacting spins in the
complex magnetic field $\overrightarrow{B}$ in Eq. (\ref{B}). The
Hamiltonian reads%
\begin{eqnarray}
H_{\text{spin}} &=&\sum_{l=1}^{d}H_{l}  \label{H_spin} \\
&=&\sum_{l=1}^{d}\overrightarrow{s_{l}}\cdot \overrightarrow{B},  \notag
\end{eqnarray}%
where $\overrightarrow{s_{l}}$ is spin-$\frac{1}{2}$ operator for the $l$th
particle of the $d$-particle ensemble. In $z$-component spin basis, the
matrix representation of $H_{\text{spin}}$\ has the form%
\begin{equation}
M_{\text{spin}}=\oplus _{l=1}^{d}M_{l},  \label{M_spin}
\end{equation}%
where%
\begin{equation}
M_{l}=\left(
\begin{array}{cc}
i\gamma & 1 \\
1 & -i\gamma%
\end{array}%
\right) .  \label{M_l}
\end{equation}%
Here $\oplus $\ denote Kronecker sum which is defined using the Kronecker
product $\otimes $ and normal matrix addition as\ $M_{l}\oplus M_{k}\equiv
M_{l}\otimes I_{2}+I_{2}\otimes M_{k}$. It is different from the direct sum
of two matrices. Equivalently, such an operation also represents the angular
momentum coupling of separate angular momenta $\overrightarrow{s_{l}}$\ and $%
\overrightarrow{s_{k}}$.

According to the graph theory \cite{D.E. Knuth}, we notice that $M_{l}\left(
\gamma =0\right) $ also represents the adjacency matrix of a two-vertex
complete graphs\textbf{\ }$k_{2}$. Meanwhile matrix$\ M_{\text{spin}}\left(
\gamma =0\right) $ is the adjacency matrix of the graph $Q_{d}=\left(
k_{2}\right) ^{\square d}$, which is constructed by Cartesian product of $n$
two-vertex complete graphs\textbf{\ }$k_{2}$. Here $G_{1}\square G_{2}$
denotes the Cartesian product of two graphs $G_{1}$ and $G_{2}$, and graph $%
Q_{d}$\ is a hypercube graph of $d$ dimensions.

On the other hand, the total spin $\overrightarrow{S}=\sum_{l=1}^{d}%
\overrightarrow{s_{l}}$\ is conservative for the Hamiltonian $H_{\text{spin}%
} $, i.e., $\left[ S^{2},H_{\text{spin}}\right] =0$. Then matrix $M_{\text{%
spin}}$\ can be diagonalized in each invariant subspace denoted by $S$. In
each invariant subspace with $S=d/2,d/2-1,...,0$ (or $1/2$), the matrix is
the same as that of Eq. (\ref{H_i})\ with $N=2S+1$\ in single-particle case.
For the case of $N=d+1$, it accords with the analysis for the engineered
chain\ in Ref. \cite{M. Christandl2}. The column in the hypercube graph
corresponds to the set of the states with the same $z$-component spin $%
S^{z}=\sum_{l=1}^{d}s_{l}^{z}$. In Fig. \ref{fig2} we take $d=4$ for an
example to illustrate our analysis.


\begin{figure*}[tbp]
\includegraphics[ bb=67 167 490 605, width=7.3 cm, clip]{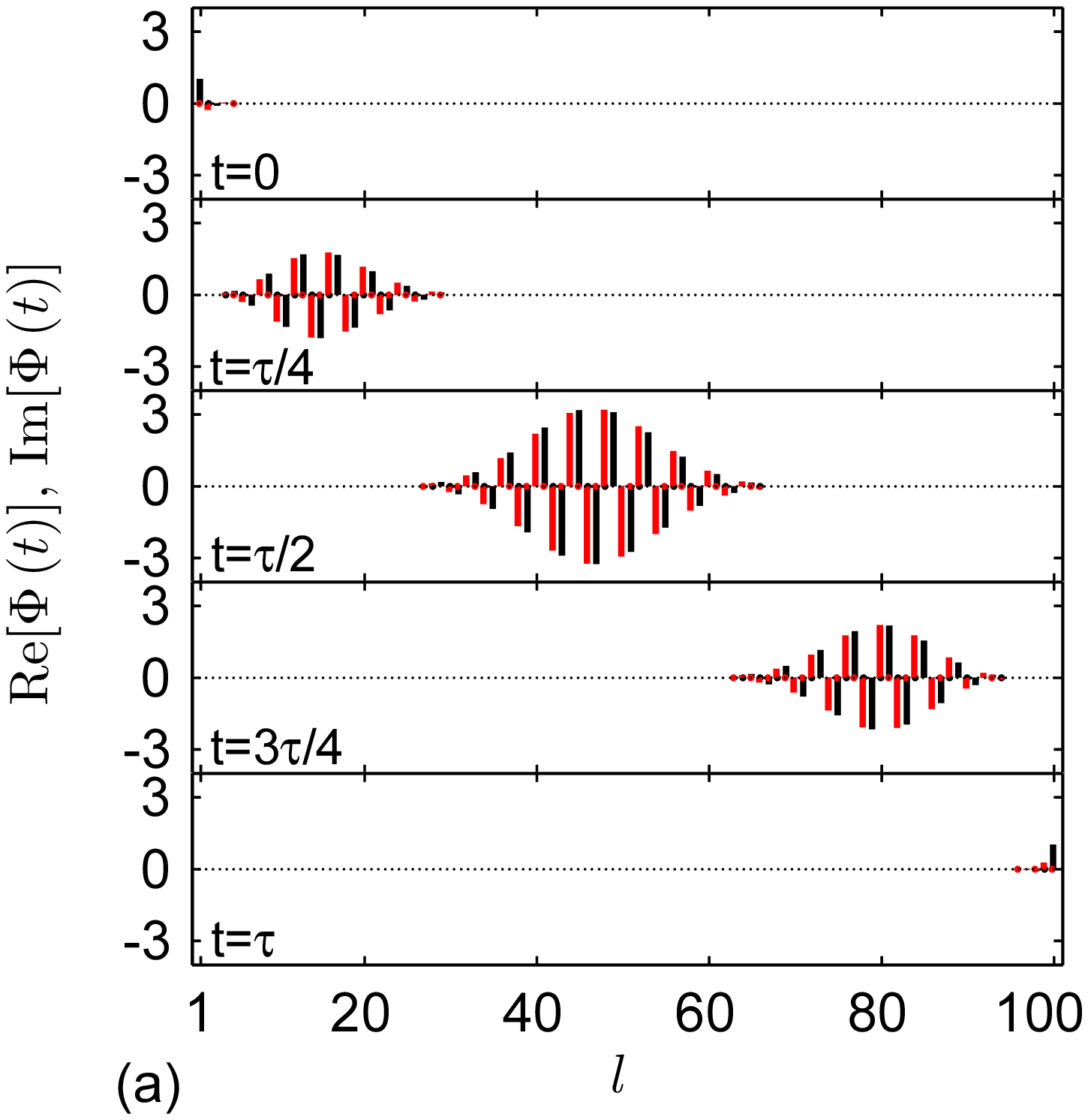} %
\includegraphics[ bb=67 167 499 605, width=7.45 cm, clip]{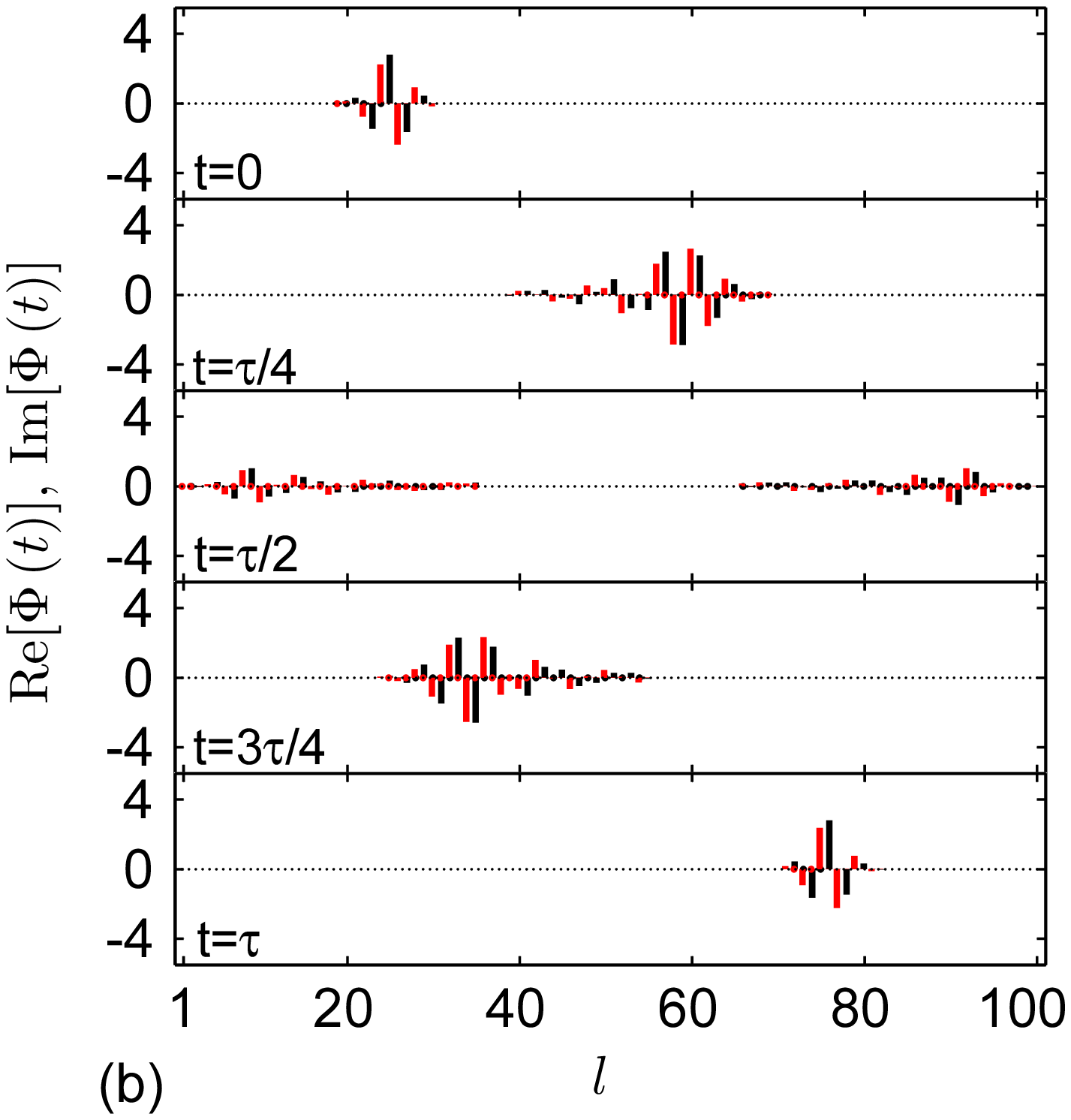} %
\includegraphics[ bb=67 167 499 605, width=7.3 cm, clip]{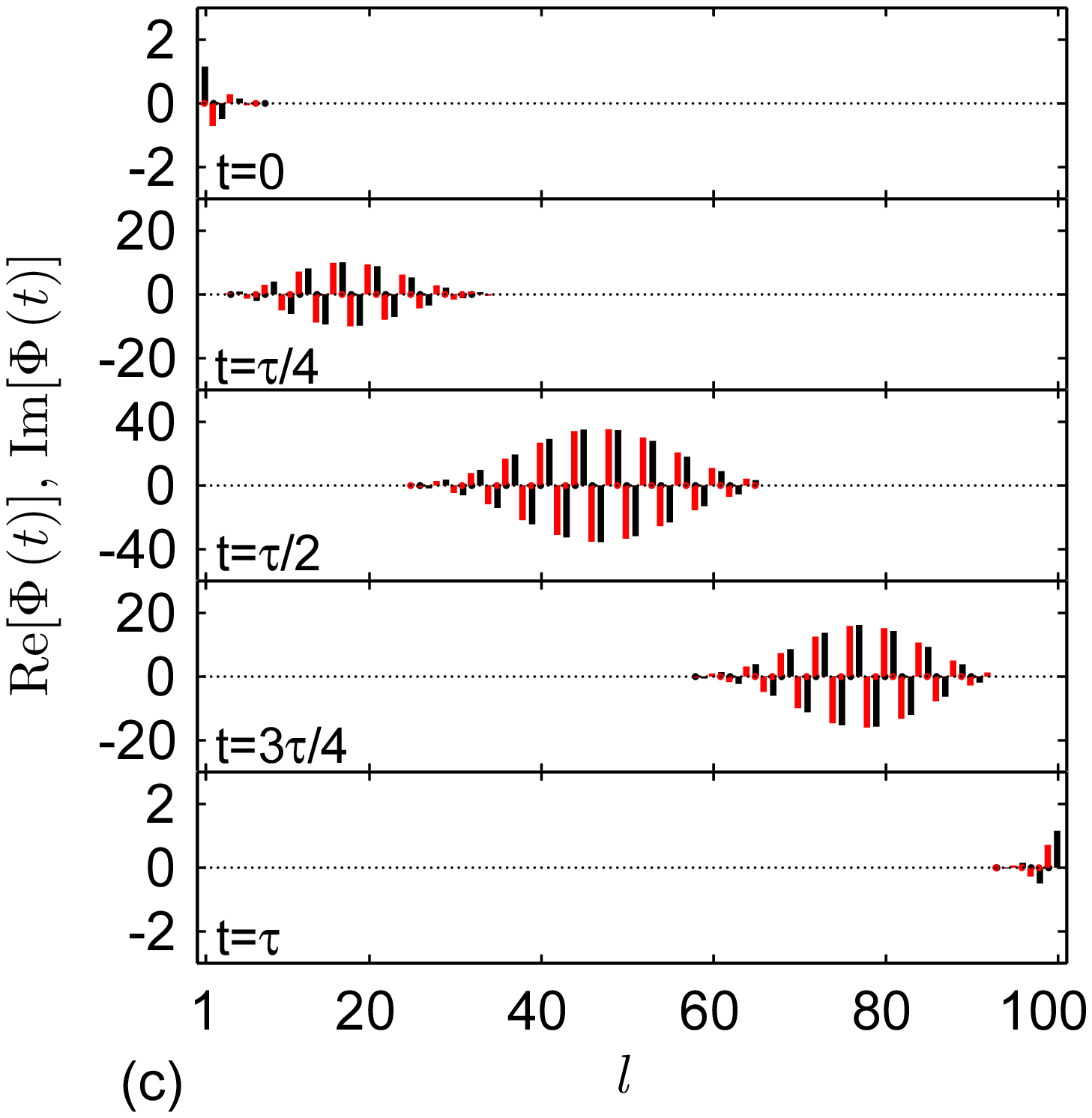} %
\includegraphics[ bb=60 167 499 605, width=7.45 cm, clip]{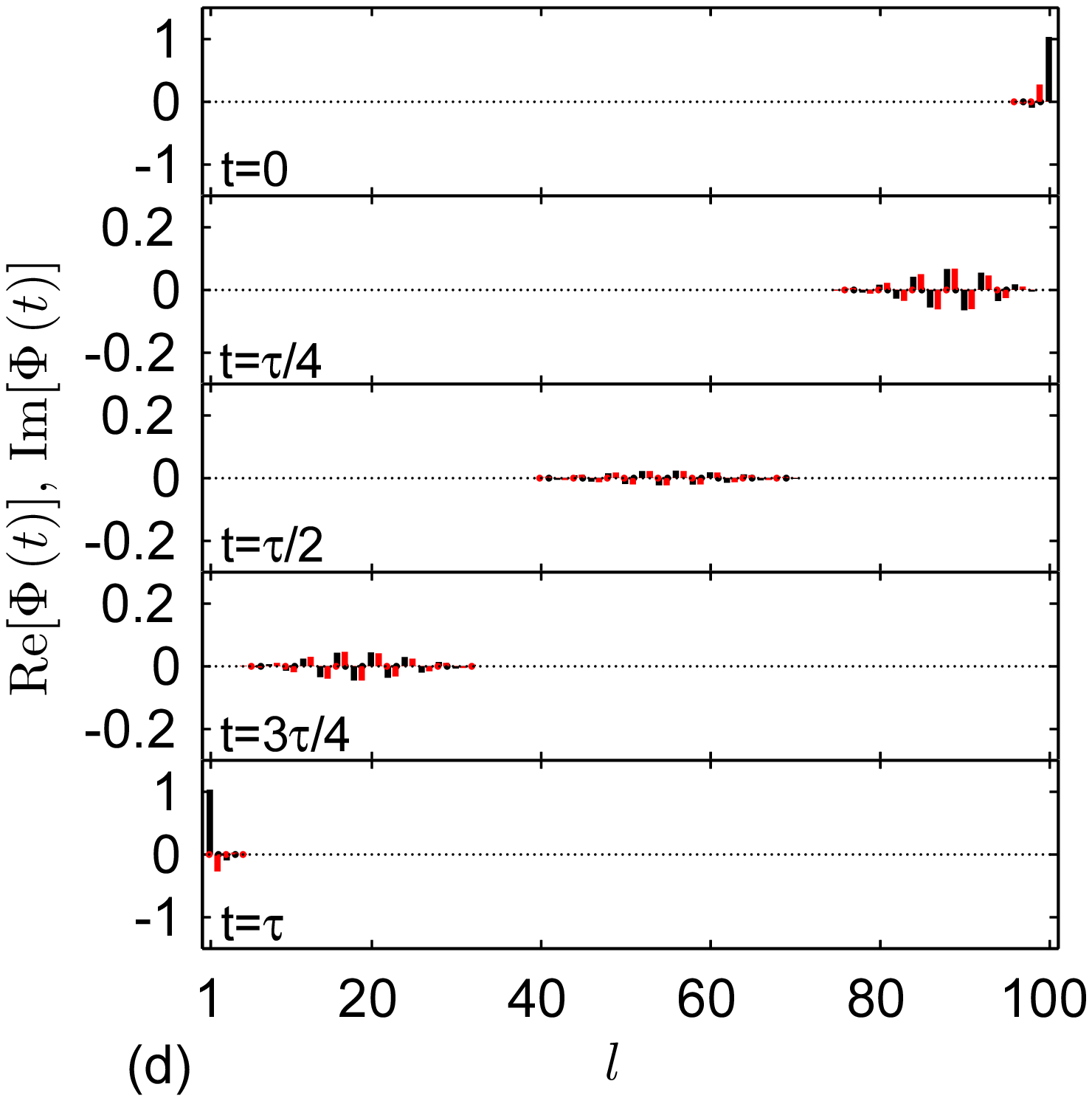}
\caption{(Color online) Plots of the time evolution for real-valued quantum
states in $N=100$\ chain. The initial state has the form of Eq. (\protect\ref%
{initial_1}). The black (red) line indicates the real (imaginary) part of
the wavefunction for (a) $\protect\gamma =0.05$, $l=1$, (b) $\protect\gamma %
=0.05$, $l=25$, (c) $\protect\gamma =0.1$, $l=1$, (d) $\protect\gamma =0.05$%
, $l=100$, respectively. It shows that the width of the initial state
becomes wider as $l$\ or $\protect\gamma $\ increases. We can see that the
final state at the time $t=\protect\tau $\ is the $\mathcal{PT}$\ function
of the initial state. In (a, b, c) the amplitudes of the evolved state
increase during the internal $\left[ 0,\protect\tau /2\right] $, then
decrease during $\left[ \protect\tau /2,\protect\tau \right] $. In contrast,
it decreases during $\left[ 0,\protect\tau /2\right] $, then increases
during $\left[ \protect\tau /2,\protect\tau \right] $ in (d). Note that the
scales of the subfigures are different.}
\label{fig3}
\end{figure*}


\section{Perfect state transfer}

\label{sec_perfectstate}Now we turn to the dynamics of the non-Hermitian
system. We notice that the imaginary linear potentials break the $\mathcal{P}
$\ symmetry but retains\ the $\mathcal{PT}$\ symmetry.

In the following, we will explain how to use such system to realize the
perfect state transfer. The time evolution of the initial state $\left\vert
\Phi \left( 0\right) \right\rangle $\ can be expressed as%
\begin{equation}
\left\vert \Phi \left( t\right) \right\rangle =e^{-i\mathcal{H}t}\left\vert
\Phi \left( 0\right) \right\rangle
=\sum\limits_{n}C_{n}e^{-iE_{n}t}\left\vert \psi _{n}\right\rangle
\label{phi_t}
\end{equation}%
where $C_{n}=\langle \phi _{n}\left\vert \Phi \left( 0\right) \right\rangle $
can be obtained from the Eq. (\ref{complete}). At time $t=\tau =\pi /\sqrt{%
\left( 1-\gamma ^{2}\right) }$, from Eqs. (\ref{ep_k}) and (\ref{PT}) we
have
\begin{eqnarray}
\left\vert \Phi \left( \tau \right) \right\rangle
&=&\sum\limits_{n}C_{n}\left( -1\right) ^{n}\left\vert \psi
_{n}\right\rangle   \label{phi_tau} \\
&=&\sum\limits_{n}C_{n}\mathcal{PT}\left\vert \psi _{n}\right\rangle   \notag
\\
&=&\mathcal{PT}\sum\limits_{n}C_{n}^{\ast }\left\vert \psi _{n}\right\rangle
.  \notag
\end{eqnarray}%
For the state with all real $C_{n}$, we have%
\begin{equation}
\left\vert \Phi \left( \tau \right) \right\rangle =\mathcal{PT}\left\vert
\Phi \left( 0\right) \right\rangle ,  \label{phi-PT}
\end{equation}%
i.e., at the instant $\tau $, the time evolution operator acts as $\mathcal{%
PT}$\ operator. For $\gamma =0$\ case, Eq. (\ref{phi-PT}) reduces to $%
\left\vert \Phi \left( \tau \right) \right\rangle =\mathcal{P}\left\vert
\Phi \left( 0\right) \right\rangle $ for arbitrary states $\left\vert \Phi
\left( 0\right) \right\rangle $. As we mentioned in the last section, Eq. (%
\ref{PT}) is not necessary for a given eigenstate. Nevertheless, Eq. (\ref%
{phi-PT}) can provide a way to construct the state to satisfy the Eq. (\ref%
{PT}). In the following we will exemplify this point and its application.

With the same mechanism of the action of operator $\mathcal{P}$, operator $%
\mathcal{PT}$ also takes the role of perfect quantum state transfer. The
flaw of this scheme is that it only applicable for some specific state.
However, if there exists local state satisfying Eq. (\ref{phi-PT}), it has a
potential for future applications.

In the following, we will show that the local state can be constructed to
perform perfect state transfer for small $\left\vert \gamma \right\vert $.
Consider a local state at the end of the chain $a_{1}^{\dagger }\left\vert
0\right\rangle $, which can be expanded with eigenstates%
\begin{eqnarray}
a_{1}^{\dagger }\left\vert 0\right\rangle  &=&\sum\limits_{m}\left\vert \psi
_{m}\right\rangle \left\langle \phi _{m}\right\vert a_{1}^{\dagger
}\left\vert 0\right\rangle  \\
&=&\sum\limits_{m}d_{m,1}^{\left[ \frac{N-1}{2}\right] }\left( \beta \right)
\left\vert \psi _{m}\right\rangle .  \notag
\end{eqnarray}%
Obviously, such a state does not satisfy Eq. (\ref{phi-PT}) and cannot be
perfectly evolved\ to the state $a_{N}^{\dagger }\left\vert 0\right\rangle $%
. Otherwise, one can construct a state satisfying Eq. (\ref{phi-PT}) based
on the state $a_{1}^{\dagger }\left\vert 0\right\rangle $\ in the way%
\begin{equation}
|\widetilde{1}\rangle =\frac{1}{\sqrt{\Omega _{1}}}\sum\limits_{m}\left\{
d_{m,1}^{\left[ \frac{N-1}{2}\right] }\left( \beta \right) +d_{m,1}^{\ast %
\left[ \frac{N-1}{2}\right] }\left( \beta \right) \right\} \left\vert \psi
_{m}\right\rangle ,  \label{1}
\end{equation}%
where $\Omega _{1}=2+\left( 1-\gamma ^{2}\right) ^{1-N}$\ is the
normalization factor. Note that the expansion coefficients of $|\widetilde{1}%
\rangle $\ are all real, so that it can evolve to the state $\mathcal{PT}|%
\widetilde{1}\rangle $. Now we will prove that such a state is local in the
case of $\left\vert \gamma \right\vert \ll 1$. Actually, rewriting the state
$|\widetilde{1}\rangle $ in the basis $\{a_{l}^{\dagger }\left\vert
0\right\rangle \}$ and neglecting the high-order terms of the Taylor
expansion, we obtain%
\begin{equation}
|\widetilde{1}\rangle =a_{1}^{\dagger }\left\vert 0\right\rangle -i\frac{%
\gamma }{2}\sqrt{\left( N-1\right) }a_{2}^{\dagger }\left\vert
0\right\rangle ,  \label{1 gamma}
\end{equation}%
which is a local state. At time $t=\tau =\pi /\sqrt{\left( 1-\gamma
^{2}\right) }$, it evolves to state\emph{\ }$\left( a_{N}^{\dagger }+i\gamma
\sqrt{\left( N-1\right) }/2a_{N-1}^{\dagger }\right) \left\vert
0\right\rangle $.

In order to demonstrate and verify the above analysis, we perform the
numerical simulation for a finite $N$-site system. The initial wave function
has the form%
\begin{equation}
|\widetilde{l}\rangle =\frac{1}{\sqrt{\Omega _{l}}}\sum\limits_{m}\left\{
d_{m,l}^{\left[ \frac{N-1}{2}\right] }\left( \beta \right) +d_{m,l}^{\ast %
\left[ \frac{N-1}{2}\right] }\left( \beta \right) \right\} \left\vert \psi
_{m}\right\rangle ,  \label{initial_1}
\end{equation}%
where $\Omega _{l}=2+2\sum\nolimits_{m=1}^{N}d_{m,l}^{\left[ \frac{N-1}{2}%
\right] }\left( \beta \right) d_{m,l}^{\ast \left[ \frac{N-1}{2}\right]
}\left( \beta \right) $\ is the normalization factor. The evolved wave
function
\begin{equation}
\Phi \left( n,t\right) =\left\langle 0\right\vert a_{n}e^{-i\mathcal{H}t}|%
\widetilde{l}\rangle   \label{phi_n_t}
\end{equation}%
is computed via exact diagonalization method. We plot the real and imaginary
parts of $\Phi \left( n,t\right) $ as function of time in Fig. \ref{fig3},
respectively. It shows that the evolution process is different from that in
a Hermitian system: the Dirac inner product of the evolved state is not
conservative. The time evolution during intervals $\left[ 0,\tau \right] $
and $\left[ \tau ,2\tau \right] $ are completely different processes.

\section{Summary and discussion}

\label{sec_summary}

In this paper, we have shown that adding the $\mathcal{PT}$-symmetric
potentials on the well-studied Hermitian quantum network \cite{M.
Christandl1,M. Christandl2}\ constructs a exactly solvable non-Hermitian
model which allows conditional perfect state transfer within the unbroken $%
\mathcal{PT}$ symmetry region, but not arbitrary. This model has
applicability and relevant to the physical situations since there exist
local states which can be transferred perfectly across long distance. In the
theoretical aspect, this work provide a nice paradigm to demonstrate the
relationship between a pseudo-Hermitian Hamiltonian and its Hermitian
counterpart in the framework of the complex quantum mechanics. On the other
hand, the simultaneity of the onset of the $\mathcal{PT}$\ symmetry breaking
for the whole eigenstates should lead to remarkable phenomena in the
dynamics of a wavepacket. This result suggests the evident observation of
the $\mathcal{PT}$\ symmetry breaking\ in optical system with complex index.

\acknowledgments We acknowledge the support of the CNSF (Grant No. 10874091
and National Basic Research Program (973 Program) of China under Grant No.
2012CB921900.

\end{document}